\documentclass[aps,prl,twocolumn,superscriptaddress]{revtex4}

\usepackage{graphicx}

\begin{document}

\newcommand{\B}{BaCuSi$_2$O$_6$ }
\newcommand{\BB}{BaCuSi$_2$O$_6$} 
\newcommand{\Sr}{Sr$_2$Cu(BO$_3$)$_2$ }
\newcommand{\Ba}{Ba$_2$Cu(BO$_3$)$_2$ }
\newcommand{\BaBa}{Ba$_2$Cu(BO$_3$)$_2$} 

\title{Role of anisotropy in the spin-dimer compound \B}

\author{Suchitra E. Sebastian}
\thanks{Current address: Cavendish Laboratory, University of Cambridge, Madingley Road, Cambridge CB3 0HE, UK}
\affiliation{Department of Applied Physics, Geballe Laboratory for
Advanced Materials, Stanford University, California 94305-4045}
\author{P. Tanedo}
\affiliation{Department of Applied Physics, Geballe Laboratory for
Advanced Materials, Stanford University, California 94305-4045}
\author{P. A. Goddard}
\thanks{Current address: Clarendon Laboratory, University of Oxford, Parks Road, Oxford OX1 3PU, UK}
\affiliation{MST-NHMFL, Los Alamos National Laboratory, Los Alamos,
NM 87545}
\author{S.-C. Lee}
\affiliation{Department of Physics, University of Florida,
Gainesville, Florida 32611}
\author{A. Wilson}
\affiliation{Department of Physics, University of Florida,
Gainesville, Florida 32611}
\author{S. Kim}
\affiliation{Department of Physics, University of Florida,
Gainesville, Florida 32611}
\author{S. Cox}
\affiliation{MST-NHMFL, Los Alamos National Laboratory, Los Alamos,
NM 87545}
\author{R. D. McDonald}
\affiliation{MST-NHMFL, Los Alamos National Laboratory, Los Alamos,
NM 87545}
\author{S. Hill}
\affiliation{Department of Physics, University of Florida,
Gainesville, Florida 32611}
\author{N. Harrison}
\affiliation{MST-NHMFL, Los Alamos National Laboratory, Los Alamos,
NM 87545}
\author{C.D. Batista}
\affiliation{MST-NHMFL, Los Alamos National Laboratory, Los Alamos,
NM 87545}
\author{I. R. Fisher}
\affiliation{Department of Applied Physics, Geballe Laboratory for
Advanced Materials, Stanford University, California 94305-4045}

\date{\today}

\begin{abstract}
We present results of magnetisation and electron paramagnetic
resonance experiments on the spin-dimer system \BB. Evidence
indicates that the origin of anisotropic terms in the spin
Hamiltonian is from magnetic dipolar interactions. Axial
symmetry-breaking is on a very small energy scale of $\leq$~11~mK,
confirming Bose Einstein condensation critical scaling over an
extended temperature range in the vicinity of the quantum critical
point.
\end{abstract}

\maketitle

A field-tuned quantum critical point (QCP) separates the low field
quantum paramagnetic phase in spin-dimer compounds from the
magnetically ordered phase in high magnetic fields. The XY
antiferromagnetic ordered state of these compounds corresponds to a
Bose-Einstein condensate (BEC) \emph{in the absence of axial
symmetry breaking}. To ascertain the extent of the BEC universal
region in each spin dimer material therefore, it is vital to
identify the nature and size of any anisotropic terms that may be
present in the Hamiltonian~\cite{Sirker1}. In order to observe BEC
critical scaling, any anisotropic terms must be small enough that
the universal region is at least a decade in reduced field.
Anisotropic terms may be directly measured by means of Electron
Paramagnetic Resonance (EPR), which has revealed for example in the
prototypical spin dimer compound TlCuCl$_3$ (triplon bandwidth
$\sim$1~meV)~\cite{Nikuni, Ruegg, Tanaka, Oosawa, Shindo}, staggered
Dzyaloshinskii-Moriya (DM) terms on the order of 0.1~meV
\cite{Glazkov}. In this report, we discuss the size and role of
anisotropy in the spin dimer compound BaCuSi$_2$O$_6$~\cite{Sasago,
Jaime, Sebastian1, Sebastian2}.

A first approximation to the equivalent magnetic lattice corresponding to the room temperature tetragonal body-centered \B lattice \cite{Sparta} is the isotropic spin Hamiltonian:

\begin{eqnarray}
{\cal{H}} = \sum_i J \mathbf{s}_{i,1}\cdot\mathbf{s}_{i,2}+
\sum_{i}\sum_{\nu}J^\prime\mathbf{s}_{i+{\hat
e}_{\alpha},\nu}\cdot\mathbf{s}_{i,\nu} \nonumber\\
+\sum_{i}J_f\mathbf{s}_{i+{\hat
e}_{\beta},1}\cdot\mathbf{s}_{i, 2} - g_\theta\mu _B\sum_{i,\nu} H
s^z_{i,\nu} \label{eq:Hamiltonian}
\end{eqnarray}

An antiferromagnetic (AF) exchange constant $J>0$ couples each pair
of Cu$^{2+}$ $s~=~\frac{1}{2}$ spins within vertical spin dimers on
Cu$_2$Si$_4$O$_{12}$ layers (where $i$ is the index of the dimer,
and $\mathbf{s}_{i1}$ and $\mathbf{s}_{i2}$ are the two spins that
form the pair $i$.) Intra-layer ($J'$) and inter-layer ($J_f$) AF
exchange constants couple nearest neighbouring spin dimers on the
lattice, indexed by ${\hat e}_{\alpha}~=~\{ {\hat x}, {\hat y} \}$,
${\hat e}_{\beta}~=~\{ {\hat z}\pm {\hat x}/2 \pm {\hat y}/2 \}$ as
defined in the high-temperature structure \cite{Ian}. For an
$s~=~\frac12$ system, energy levels corresponding to the isotropic
Hamiltonian are a groundstate singlet, and three degenerate triplet
excited states, separated from the groundstate by the spin gap
$\Delta$. An applied magnetic field $H$ introduces a Zeeman
splitting term in the Hamiltonian, scaled by the $g$-factor
$g_\theta$, corresponding to the orientation of $H$. Experimental
results reveal a critical magnetic field $H_{c1}~\sim~$23.5~T at
which the system orders for $H~\|~c$, and comparison of the measured
phase boundary with the model Hamiltonian in
Eqn.~\ref{eq:Hamiltonian} gives values of $J$ = 4.45 meV, $J^\prime$
= 0.51 meV, and an even smaller value of $J_f$~\cite{Jaime,
Sebastian1}.

The above description neglects anisotropic exchange interactions
which can reduce the symmetry of the ordered state. In the more
general spin Hamiltonian, the bilinear spin term $\textbf{s}_i \cdot
\textbf{s}_j~\equiv~\textbf{s}_i~\emph{\textbf{I}}~\textbf{s}_j$ in
Eqn.~\ref{eq:Hamiltonian} is replaced by the exchange interaction
$\sum_{i,j=1}^3 \textbf{s}_i\tensor{T}_{i,j}\textbf{s}_j$, where
\begin{eqnarray}
\tensor{T}_{i,j}~&=&~3T_s~+~\tensor{T}_{as}~+~\tensor{T}_{sm},\\
\tensor{T}_{as}~&=&~\sum\frac12(T_{i,j}-T_{j,i})(\mathbf{e_i}~\times~\mathbf{e_j}),\label{asymm}\\
\tensor{T}_{sm}~&=&\frac12(T_{i,j}+T_{j,i})_{i\neq j}\label{symm}
\end{eqnarray}
$T_s$ is a scalar, while the antisymmetric term $\tensor{T}_{as}$ of
the form $\mathbf{D} \cdot [\mathbf{s}_i \times \mathbf{s}_j]$, and
the traceless symmetric term $\tensor{T}_{sm}$ of the form
$\mathbf{s}_i\tensor{\Gamma}\mathbf{s}_j$ mix spin
components~\cite{Moriya}. The DM interaction arising from the
spin-orbit coupling is antisymmetric in nature, leading to
singlet-triplet ($ST$) mixing to the lowest order of perturbation.
Magnetic dipolar interactions, however, belong to the traceless
symmetric category. Terms of this nature lead to intra-triplet
matrix elements entering to the lowest order of perturbation, but
$ST$ mixing to higher orders of perturbation only for $H$ oriented
away from the dipole vector. An additional possible source of
anisotropic exchange interactions is the reduced symmetry of the \B
lattice at low temperatures (recent structural analysis has revealed
a weak orthorhombic structural transition at $\sim$100~K,
accompanied by an incommensurate lattice modulation \cite{Samulon}).

An applied magnetic field with a component perpendicular to either
the DM or dipolar vector would result in spin non-conservation due
to mixing of singlet and triplet energy levels with inequivalent
$S_z$. The symmetry of the Hamiltonian would therefore reduce from
rotational invariance (U(1)) to the discrete Z$_2$ group, resulting
in a magnon spectrum in the ordered phase with a finite gap to the
minimum, rather than a gapless Goldstone mode. The extent of
symmetry breaking, however depends on the order of perturbation at
which spin non-conserving terms enter the spin Hamiltonian; it is
therefore important to identify the origin of any anisotropic
interaction in the Hamiltonian. Previous measurements of a BEC
critical exponent in \B down to 0.03 K \cite{Sebastian1, Sebastian2}
provide empirical evidence for the absence of U(1) symmetry-breaking
terms down to this energy scale for $H\|c$. In this rapid
communication, we discuss more direct experimental evidence that
estimates the size of any axial symmetry-breaking due to anisotropic
exchange terms in the spin Hamiltonian.

Single crystals were grown using the slow cooling flux technique.
Polycrystalline \B precursor was synthesized by a solid state
reaction of BaCO$_3$, SiO$_2$, and CuO between temperatures of 900
and 1050$^\circ$C in flowing oxygen, with repeated regrinding.
Single crystals up to 1 g were grown by heating a 2:1 molar mixture
of ground polycrystalline material and LiBO$_2$ flux in a platinum
crucible to 1000$^\circ$C in air, followed by slow cooling to
875$^\circ$ and decanting by centrifuge.

\begin{figure}[htbp]
\includegraphics[width=0.45\textwidth]{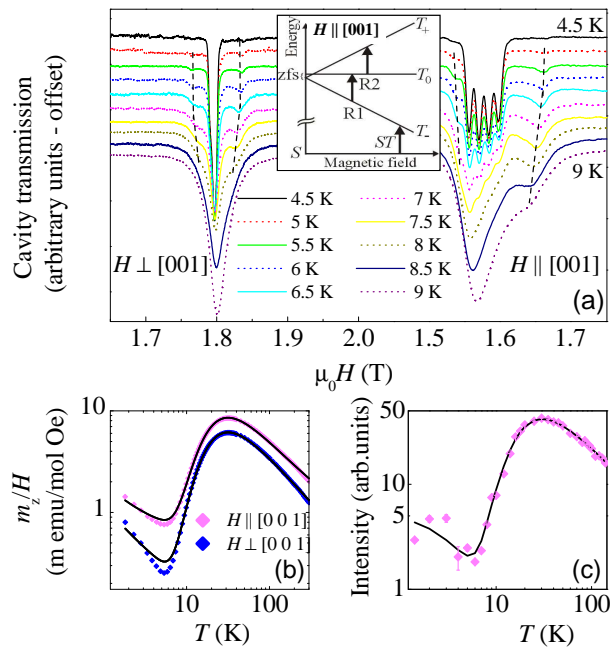}
\centering \caption {[Colour online] (a) EPR spectrum measured at
different $T$ for $H\|c$ and $H\perp c$ using a resonant frequency
of 51.8~GHz. For clarity, data for different $T$ have been offset.
Weak shoulders appear and split from the central EPR peak upon
reducing the temperature (see dashed lines and main text for
explanation). The central peak also exhibits hyperfine splitting for
$H\parallel c$ (see main text). The inset shows a schematic of the
possible transitions between different spin states. (b) temperature
dependence of the magnetic susceptibility for a field of 5000~Oe
aligned parallel and perpendicular to the crystalline $c$-axis. (c)
measured temperature dependence of the integrated EPR intensity for
$H\| c$. Solid lines in panels (b) and (c) show fit to isolated
dimer model including a Curie term to account for paramagnetic
impurities.} \label{peaksplit}
\end{figure}

EPR measurements were performed for a range of frequencies between
26~GHz and 660~GHz for both $H\parallel c$ and $H\perp c$. At
elevated temperatures, population of the Zeeman split triplet states
results in intra-triplet transitions, observed as sharp dips in the
transmission through a microwave cavity (Fig.~\ref{peaksplit}a). At
the highest temperatures, a single EPR peak is observed,
corresponding to the degenerate transitions between triplet levels
(in the absence of anisotropic terms). The intensity (integrated
area under the peak) diminishes rapidly upon reducing the
temperature below $\sim 30$~K (Fig.~\ref{peaksplit}c), as the
triplet states depopulate. A residual intensity can be observed to
persist, and even grow, at the lowest temperatures, exhibiting
hyperfine splitting (into 4 distinct peaks). The 4-peak pattern is
consistent with hyperfine coupling associated with a localized
$s~=~\frac12$ and a lone Cu nucleus ($I~=~\frac32$). From fits to
the temperature dependence of the total integrated intensity
(Fig.~\ref{peaksplit}c), we conclude that the central portion of the
spectrum observed below about 5~K is dominated by a small
concentration ($0.5(1)\%$) of isolated paramagnetic Cu$^{2+}$
defects ($s~=~\frac12$) in agreement with estimates from DC magnetic
susceptibility ($\chi = m_z/H$) measurements of the same sample made
using a SQUID magnetometer (Fig.~\ref{peaksplit}b). Since the
hyperfine splitting has the same $g$ factor and anisotropy as the
lattice, the associated Cu$^{2+}$ defects likely correspond to
singly-occupied dimers. Perhaps the most important feature in the
data is the appearance of distinct, albeit weak shoulders which
split from the central peak at low temperatures (dashed lines in
Fig.~\ref{peaksplit}a). The vanishing intensities of these peaks as
$T~\rightarrow~0$ indicate that they involve EPR transitions within
the triplet state.

\begin{figure}[htbp]
\includegraphics[width=0.4\textwidth]{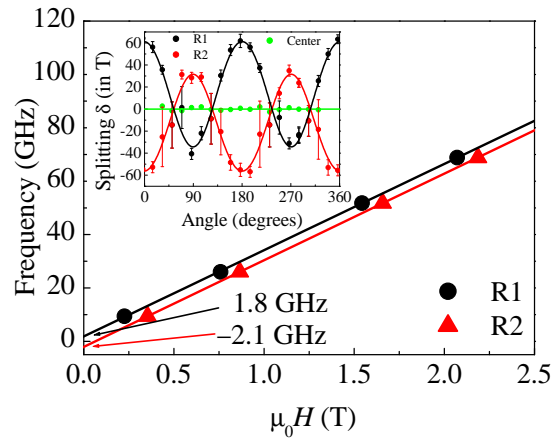}
\centering \caption {[Colour online] Magnetic field dependence of
experimentally measured intra-triplet splitting (labelled R1 and R2
following Fig.~\ref{peaksplit}) for $H\| c$. The solid lines are
linear fits to the intra-triplet splitting as a function of magnetic
field, extrapolated down to zero magnetic field. The inset shows the
angular dependence of this splitting measured using a resonant
frequency of 51.8~GHz at a temperature of 6~K. The solid lines show
a fit to Eqn.~\ref{angledep} with $D$~=~0.091 K.} \label{zfs}
\end{figure}

The frequency and field orientation dependences of the split
shoulders are displayed in Fig.~\ref{zfs}. The splitting suggests an
anisotropic zero-field-splitting (zfs) interaction in the spin
Hamiltonian, as confirmed from the frequency dependent measurements
for $H\parallel c$, i.e the splitting ($\delta$) is
field-independent. The field orientation dependence of the splitting
can be fit by the equation:

\begin{equation}
\delta = \frac12 D (3\cos^2\theta~-~1)/\mu_B g_\theta
\label{angledep}
\end{equation}
where $\theta$ is the angle between $H$ and the vector between the
intra-dimer Cu$^{2+}$ sites (i.e. the $c$-axis). One possible source
of such an angular dependence is a symmetric anisotropic interaction
(Eqn.~\ref{symm}) in the Hamiltonian of the form $D\hat{S}_z^2$,
where $S$ is the total spin (=~1) of the dimer and $D=\Gamma_{zz}$.
The value of the anisotropic parameter $D$ is found from the angle
dependence to be 0.091(3)~K. Similar shoulders have been observed in
Ref.~\cite{Zvyagin}, and the angle dependence of the shoulders fit
to $D$~$\sim$0.1~K. A more direct measure of the anisotropic
parameter involves a determination of the zfs of triplet levels by
extrapolating the frequency dependence to zero field
(Fig.~\ref{zfs}). By this method, zfs of 1.8(3)~GHz [0.09(1)~K] and
2.1(1)~GHz [0.10(1)~K] are obtained, in agreement with the value
from the angle dependence. We note also that fits to the central
high temperature peak yield extremum values for the g-factor of
$g_\parallel = 2.307(3)$ and $g_\perp = 2.057(3)$, in excellent
agreement with susceptibility and X-band EPR data~\cite{Zvyagin}.

\begin{figure}[htbp]
\centering
\includegraphics[width=0.35\textwidth]{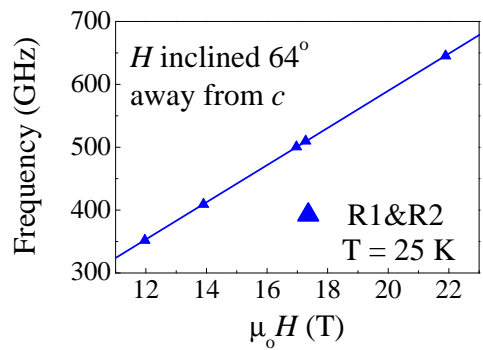}
\caption {High field linear field dependence of the intra-triplet
Zeeman splitting with $H$ inclined with respect to [001].}
\label{linearesr}
\end{figure}

The observed zfs is consistent with either an antisymmetric DM
interaction (Eqn.~\ref{asymm}), with $\mathbf{D}~\|~c$~\cite{DM} and
$|\mathbf{D}|=4.8(2)$~K or a symmetric dipolar intra-dimer
interaction. A DM interaction would mix singlet ($S$) and triplet
$\hat{S}_z = 0$ ($T_0$) states for $H\|c$ to lowest order - which
although not U(1) symmetry-breaking, would allow a significant EPR
transition between the $S$ and triplet $\hat{S}_z = -1$ ($T_{-}$)
states (labelled $ST$ in Fig.~\ref{peaksplit}). Simulations which
include an antisymmetric anisotropy predict similar intensities for
the triplet transition (labelled R1 and R2 in Fig.~\ref{peaksplit})
and the high-field $ST$ transition at 6~K. In contrast, for $H\|c$,
an $ST$ EPR transition is forbidden by symmetry for an intra-dimer
dipolar interaction, and would be negligibly small due to higher
order mixing of the $S$ and $T_-$ states for an inter-dimer dipolar
interaction. For $H\perp c$, the extent of mixing of the $S$ and
$T_-$ states (which would break $U(1)$ symmetry) would be
considerable for a first order DM process, but very weak for higher
order intra- and inter-dimer dipolar processes. Therefore, the DM
interaction would lead to significant non-linear field-dependence of
the intra-triplet transition frequency for fields approaching
$H_{c1}$, whereas the dipolar interaction would not. High-field EPR
measurements reveal no sign of $ST$ transitions in the vicinity of
$H_{c1}$, or any unusual behaviour of the intra-triplet EPR
transitions for $H\perp c$ up to 21.7~T (Fig.~\ref{linearesr}),
strongly indicating that the observed low-temperature EPR splitting
arises from dipolar interactions which are symmetric in nature.
Furthermore, as we describe below, both the magnitude of the zero
field splitting and its $T$-dependence are consistent with dipolar,
and not DM interactions.

The dilute concentration of well separated thermally activated
triplets at low $T$, $H$ provides the ideal conditions to observe
the effects of magnetic dipolar interactions in \B. The magnetic
lattice comprises closely spaced pairs of Cu$^{2+}$ ions
($r_\parallel=2.74$~\AA~intra-dimer spacing at room temperature)
which are well separated from each other ($r_\perp=7$~\AA~between
dimers). The zero-field dipolar splitting of triplets is given by
$\frac{\mu_0}{16 \pi r_{\parallel}^3}(2
g_\parallel^2+g_\perp^2)\mu_B^2$, which has the value 0.113~K for
$r_{\parallel}$~=~2.74~{\AA}. This is remarkably close to the
measured zfs~$\sim$~0.1~K, suggesting that the origin is indeed
dipolar interactions. The reason for the collapse of the dipolar
splitting at higher temperatures can be understood in terms of
`exchange narrowing'. At low $T$, the triplets are dilute and long
lived on EPR time scales ($\sim1/f$). The zfs results from the
anisotropic dipolar field that each spin within a dimer experiences
due to its pair [$\propto
3(s_1\cdot\hat{z})(s_2\cdot\hat{z})-(s_1\cdot s_2)$]. However, as
$T$ is raised, and more triplets are excited, any given Cu spin will
experience strong fluctuations in the local dipolar fields due to
the exchange-induced co-flipping between neighboring dimers of
opposing spin projection. As more triplets are excited, such
co-flipping leads to faster fluctuations of the local dipolar fields
until they are eventually averaged out on EPR time scales, and the
dipolar splitting vanishes. In fact, most of the linewidth observed
in these experiments can be attributed to nuclear and dipolar
spin-spin interactions (both intra- and inter-dimer). As $T$ is
raised, exchange averaging leads to a gradual reduction in the
second moment of the dipolar field distribution, and to a narrowing
of the spectrum. We do indeed observe a further narrowing of the
spectrum at higher $T$ (not shown), which has also been observed in
Ref.~\cite{Zvyagin}.

\begin{figure}[htbp]
\centering
\includegraphics[width=0.47\textwidth]{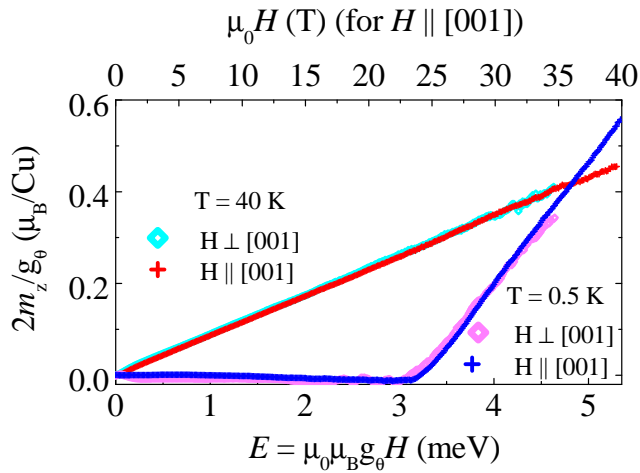}
\caption{[Colour online] Magnetisation measured as a function of $H$
up to 40~T. The applied field $H\|c$ and $H\perp c$ ($g_\|$~=~2.31,
$g_\perp$~=~2.05) is represented as an equivalent energy
$E~=~\mu_0\mu_Bg_\theta H$~meV on the lower $x$-axis, such that both
orientations share the same axis. The upper $x$-axis shows $H$ along
the $c$-axis.} \label{mag}
\end{figure}

We also discuss the directional anisotropy in the value of $g$,
which arises from crystal electric field splitting of the Cu$^{2+}$
energy-levels in a tetragonal environment, and spin orbit
interactions. The anisotropy in magnetic susceptibility for $H\perp
c$ and $H\| c$ (shown in Figure \ref{peaksplit}b) reflects an
anisotropy in $g$-values. Fits to an isolated dimer model using the
Bleaney-Bowers equation~\cite{Bleaney} yield a value of
$J$~=~4.40(2)~meV and values of $g_\perp$~=~2.03(5) and
$g_\|$~=~2.31(5) in agreement with EPR results, where $\parallel$
and $\perp$ refer to $H$ oriented parallel and perpendicular to the
$c$-axis. In addition to scaling the low field susceptibility, the
effect of this $g$-anisotropy is to scale the value of $H_{c1}$ with
$g_\theta$.

High field magnetisation measurements were performed in pulsed
magnetic fields of 500~ms pulse duration up to 40~T. Data were
obtained using a wire-wound sample extraction magnetometer in which
the sample is inserted or removed from the detection coils \emph{in
situ}. Figure \ref{mag} shows the uniform magnetisation ($m_z$) of
\B as a function of field for $H\|c$ and $H\perp c$. The value of
$m_z$ for both directions of $H$ has been normalised by 2/$g_\theta$
at 40~K so that the saturation value is 1~$\mu_B$/Cu. The upturn in
$m_z$ measured at 0.5~K occurs at $H_{c1}$, at which magnetic
ordering takes place. Linear extrapolation of $m_z$ is used to
determine the value of $\Delta=g_\theta\mu_BH_{c1}$ for $H\|c$ and
$H\perp c$. The value of spin gap thus extracted is
$\Delta$~=~3.2(1)~meV for $H\|c$ and $H\perp c$. As expected, the
values of $H_{c1}$ and $m_z$ scale with $g_\theta$ for $H$ along
different crystal axes. The effect of the anisotropy in $g$ appears
only in the Zeeman term in~Eqn.\ref{eq:Hamiltonian}, indicating that
the $g$-tensor is diagonal, and does not lead to mixing of the spin
components.

Having carefully examined any possible source of anisotropies in the
Hamiltonian describing \BB, we conclude that the dominant effect
arises from symmetric dipolar interactions, which lead to U(1)
symmetry-breaking only to second order. The magnitude of U(1)
symmetry-breaking due to the intra-dimer dipolar interaction is of
the order of $\sim |D|J'/J$ = 11 mK only for $H\perp c$, whilst of
the order of $\sim (\frac{r_\parallel}{r_\perp})^3|D|J'/J$ = 0.7 mK
for all $H$ due to inter-dimer dipolar interaction. Indeed, the
family of spin-dimer compounds is distinct from other XY
antiferromagnets due to the small size of $J'/J$ and
$r_\parallel/r_\perp$, which `protect' the U(1) symmetry down to low
energy scales in the case of symmetric anisotropies. EPR
measurements on \B confirm that the correspondence drawn with the
BEC universality class is justified in a significant region of
criticality near the QCP.

We acknowledge discussions with M. Jaime, S. Zvyagin, R. Stern, and
N. Dalal and experimental assistance from E. Samulon. This work is
supported by the NSF Grants No. DMR-0134613, No. DMR-0239481 and No. DMR-0645461. Experiments performed at the NHMFL were supported by NSF, Florida State, and DOE. I. R. F. acknowledges support from the Alfred P. Sloan
Foundation and S. E. S. from the Mustard Seed Foundation $\&$
Trinity College, Cambridge.


\begin{thebibliography}{1}
\bibitem{Ian}For simplicity, the unit vectors in the Hamiltonian refer to the high $T$ structure with I4/mmm symmetry, in which notation the nearest neighbour inter-dimer coupling lies along the lattice vectors. On cooling below 610~K, the room $T$ structure is modified to I4$_1$/acd~\cite{Sparta} in which the in-plane lattice vectors are rotated by 45$^\circ$.
\bibitem{DM}Since the low $T$ orthorhombic distortion of the I4$_1$/acd space
group is subtle~\cite{Samulon}, the dimer axis is anticipated to
closely resemble an axis of 4-fold symmetry, and any finite
$\mathbf{D}$ would lie close to the dimer axis.
\end{thebibliography}
\end{document}